\documentstyle[prb,aps,multicol,epsf]{revtex}
\renewcommand{\narrowtext}{\begin{multicols}{2} \global\columnwidth20.5pc}
\renewcommand{\widetext}{\end{multicols} \global\columnwidth42.5pc}
\begin{document}
\draft
\title {Magnetic Properties of Yb$_2$Mo$_2$O$_7$ and Gd$_2$Mo$_2$O$_7$ \\ 
from rare earth M\"ossbauer Measurements}

\author{J. A. Hodges, P.Bonville, A. Forget}
\address{Commissariat \`a l'Energie Atomique, Centre d'Etudes de Saclay \\
Service de Physique de l'Etat Condens\'e, 91191 Gif-sur-Yvette, France}
\author{J. P. Sanchez, P. Vulliet}
\address{Commissariat \`a l'Energie Atomique, Centre d'Etudes
de Grenoble \\ Service de Physique Statistique, Supraconductivit\'e et
Magn\'etisme, 38054 Grenoble, France}
\author{M. Rams, K. Kr\'olas}
\address{Institute of Physics, Jagiellonian University, 30-059 Krak\'ow, 
Poland}
 
\maketitle

\begin{abstract}

Using $^{170}$Yb and $^{155}$Gd M\"ossbauer measurements down to 
$\sim$ 0.03\,K, we have examined the semiconducting pyrochlore
Yb$_2$Mo$_2$O$_7$ where the Mo intra-sublattice interaction is 
anti-ferromagnetic
and the metallic pyrochlore Gd$_2$Mo$_2$O$_7$ where this interaction is
ferromagnetic. Additional information was obtained from susceptibility, 
magnetisation and $^{172}$Yb perturbed angular correlation measurements.
The microscopic measurements evidence lattice disorder  which is important in
Yb$_2$Mo$_2$O$_7$ and modest in Gd$_2$Mo$_2$O$_7$. Magnetic
irreversibilities occur at 17\,K in Yb$_2$Mo$_2$O$_7$ and at 75\,K in
Gd$_2$Mo$_2$O$_7$ and below these temperatures the rare earths carry magnetic
moments which are induced through couplings with the Mo sublattice.
In Gd$_2$Mo$_2$O$_7$, we observe the steady state Gd hyperfine populations at 
0.027\,K are out of thermal equilibrium, indicating that Gd and Mo spin 
fluctuations persist at very low temperatures. Frustration is thus operative 
in this essentially isotropic pyrochlore where the dominant Mo 
intra-sublattice interaction is ferromagnetic.
\vskip 0.5cm

\end{abstract}

\pacs{PACS numbers: 76.80+y, 75.50.Lk, 75.30-m }

\narrowtext

\section {Introduction}

In the rare earth pyrochlores R$_2$T$_2$O$_7$ (R is a trivalent rare earth, 
T a tetravalent transition or sp metal ion), each of the two cationic 
sublattices forms a network of corner sharing tetrahedra
\cite{subramanian93,greedanLB92}. For this arrangement, each sublattice is
prone to geometrically derived magnetic frustration
\cite{diep94,moessner01,waterloo01}, with the behaviour in each particular 
compound depending on the sign, size and anisotropy of the interionic 
couplings.
The main signatures of magnetic frustration are the absence of long range 
order and the persistence of dynamic short range magnetic
correlations as $T \to 0$ 
\,\cite{villain}.

The rare earth molybdates R$_2$Mo$_2$O$_7$ may be formed with the rare 
earths from Nd$^{3+}$ to Lu$^{3+}$
\,\cite{subramanian93}.
For R from Nd to Gd, they show metallic-like behaviour,
whereas for R from Dy to Lu (and with Y), they show semiconducting behaviour 
with an activation energy of the order of 15 meV
\,\cite{greedanLB92,greedan87,ali89}.
The change in behaviour (``metal-semiconductor crossover'') is chiefly linked 
with changes within the Mo-O subsystem and it has been related to variations 
in the Mo-O bond lengths associated with the changing lattice parameter 
(lanthanide contraction)
\cite{katsufuji00} and in the Mo-O-Mo bond angles
\cite{moritomo01}.
The metal-semiconductor crossover also leads to profound changes in the 
magnetic properties. 
In the semiconducting compounds, the Mo intra-sublattice coupling is 
antiferromagnetic 
\cite{greedan86},
whereas in the metallic compounds, it is ferromagnetic
\cite{ali89}.

In the R$_2$Mo$_2$O$_7$, 
the dominant exchange interaction is that within 
the d-ion sublattice. The next most important interaction is the 
inter-sublattice exchange with the interaction within the f-ion sublattice 
the weakest of the three. For the d-ion sublattice, 
the dominant exchange mechanism is different either side of the 
metal-semiconductor 
crossover. In the semiconducting compounds, antiferromagnetic superexchange 
prevails whereas in the metallic compounds, the direct ferromagnetic coupling 
involving the Mo 4d-spin density at the Fermi level dominates \cite{kang02}.
An important feature controlling the properties of most of the R$^{3+}$ ions
and the way they are magnetised through coupling with the Mo$^{4+}$ 
sublattice is
the crystal field interaction, which fashions the wave functions of the 
R$^{3+}$ ground state. This consideration does not concern the S-state 
Gd$^{3+}$ ion which is essentially immune to the influence of crystal fields.

In Y$_2$Mo$_2$O$_7$ , as in the other semiconducting
molybdates, the Mo-Mo interaction is antiferromagnetic. 
A magnetic irreversibility occurs at 22\,K evidencing a spin-glass-like 
transition
\cite{greedan86}. This was a surprising result for a compound which 
seemed to be crystallographically ordered.
Recent structural studies have shown, however, that the Mo-Mo 
(and some of the Y-O) bond lengths are disordered 
\cite{booth00}, raising the possibility that the spin-glass-like transition 
is linked to the disorder and frustration. 
$^{89}$Y nuclear magnetic resonance measurements have confirmed the presence 
of lattice disorder
\cite{keren01}.
Magnetic frustration is clearly operative in Y$_2$Mo$_2$O$_7$:
the spin fluctuations associated with the short range correlated Mo spins do
not die out as T $\to$ 0, but rather tend to a temperature independent rate
\cite{dunsiger96} and neutron scattering measurements have shown that
the spin-glass-like transition occurs mainly in the frequency domain
\cite{gardner99}.
In Gd$_2$Mo$_2$O$_7$, as in the other metallic molybdates, the Mo-Mo
interaction is ferromagnetic. A transition occurs with a temperature that 
depends on the sample ($T_C \simeq 55$\,K in Refs.\onlinecite{ali89,raju92} 
and $\simeq$ 65\,K in Ref.\onlinecite{katsufuji00}), and it is accompanied by 
a magnetic irreversibility. 
Little information is available concerning the possible influence of 
frustration in this compound
although specific heat measurements down to $\sim$ 1.8\,K have suggested that 
the low energy tail of the density of magnetic excitations does not tend to 
zero 
\cite{raju92}.

We present a study of  semiconducting
Yb$_2$Mo$_2$O$_7$ and metallic Gd$_2$Mo$_2$O$_7$ based chiefly on rare earth
M\"ossbauer measurements (respectively using the isotopes $^{170}$Yb and
$^{155}$Gd) down to $\sim$ 30\,mK. Magnetic measurements were also made for 
the two compounds, and $^{172}$Yb perturbed angular 
correlation (PAC) were carried out for Yb$_2$Mo$_2$O$_7$ up to $\sim$ 1000\,K.
We obtain 
information concerning the static and dynamic magnetic properties of the rare 
earths and concerning the ordering and spin-dynamics of the Mo$^{4+}$ 
sub-lattice and its coupling with the R$^{3+}$ sublattices. Making use of the
microscopic nature of the M\"ossbauer probes, we discuss the information
obtained concerning local symmetry lowering and bond disorder in the two
compounds, which are {\it a priori} expected to be crystallographically
ordered.

After outlining some background properties of the two samples in section 
\ref{sectionsamples}, we present the results concerning Yb$_2$Mo$_2$O$_7$
in section \ref{sectionyb} and concerning Gd$_2$Mo$_2$O$_7$ in section
\ref{sectiongd}. Section \ref{sectiondisc} contains the summary and 
discussion.

\section {Samples and background rare earth single ion properties.}
\label{sectionsamples}

The polycrystalline samples of  Yb$_2$Mo$_2$O$_7$ and Gd$_2$Mo$_2$O$_7$ were
prepared by first reacting Mo and MoO$_3$ in a sealed container to form
MoO$_2$ which was then reacted under argon with the R$_2$O$_3$.
Room temperature X-ray diffraction measurements show both samples are
single phase and provide the cubic lattice parameter:
10.146\AA\ (Yb$_2$Mo$_2$O$_7$) and 10.366\AA\ (Gd$_2$Mo$_2$O$_7$). 

The cubic pyrochlore structure (space group Fd$\bar 3$m) contains eight 
formula units per unit cell. The R$^{3+}$ ions situated  at the 16b sites 
(point symmetry: $\bar 3$m), form a network of corner sharing tetrahedra. Each
of the four R$^{3+}$ making up a tetrahedron has its local symmetry axis along
one of the [111] directions. The Mo$^{4+}$ situated at the 16d sites 
(point symmetry: $\bar 3$m), also form a network of corner sharing tetrahedra 
which is displaced by (1/2,1/2,1/2) relative to that of the R$^{3+}$.

Yb$^{3+}$ (4f$^{13}$) has 8 sublevels in its spin-orbit derived ground state
($^2$F$_{7/2}$). The degeneracy is partially
lifted by the crystal electric field to 
leave four Kramers doublets. In an isomorphous pyrochlore (Yb$_2$Ti$_2$O$_7$),
the total energy separation of the four doublets is about 1000\,K and 
the well isolated ground state Kramers doublet has planar anisotropy 
with the easy magnetisation plane lying perpendicular to the appropriate 
local [111] direction 
\cite{hodges01}.

Gd$^{3+}$ (4f$^7$) is an S-state ion ($^8$S$_{7/2}$). Both its anisotropy and
the amount of crystal field degeneracy lifting within the 8 
sublevels of the ground state are quite small. 
In the context of the present study, this ground 
state can be taken to be an initially degenerate S=7/2 state, whose
degeneracy is lifted only by a molecular or applied magnetic field.

\section {Yb$_2$Mo$_2$O$_7$}
\label{sectionyb}

\subsection {Susceptibility and magnetisation measurements.}
\label{secsusc}

\begin{figure}
\epsfxsize=7 cm
\centerline{\epsfbox{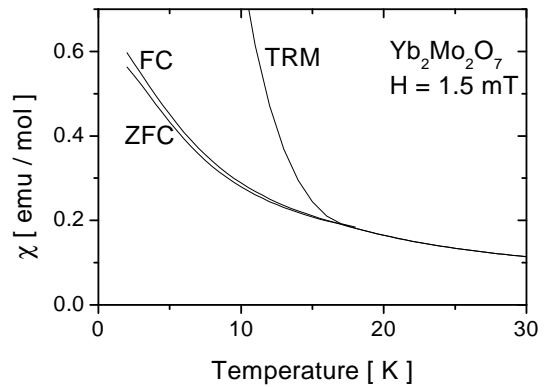}}
\vspace {0.5cm}
\caption{Magnetic susceptibility of Yb$_2$Mo$_2$O$_7$ 
with an applied field of 1.5\,mT in the field cooled (FC), zero field 
cooled (ZFC) and thermo-remanent magnetisation (TRM) configurations. The use 
of the TRM dependence (which was scaled to fit the figure) provides a more 
accurate assessment of the irreversibility temperature (17\,K).} 
\label{figybsusc}
\end{figure}

The magnetic susceptibility of Yb$_2$Mo$_2$O$_7$ was measured from 300 
to 2\,K. Both the field cooled (FC) and zero field cooled (ZFC) variations 
change monotonically as the temperature is lowered and there is a small 
irreversibility with an onset in the range 16 - 18\,K (Fig.\ref{figybsusc}).

To obtain a more accurate value for the irreversibility temperature 
$T_{irr}$, we measured the thermo-remanent magnetisation (TRM) after field 
cooling with 5\,T down to 2\,K (Fig.\ref{figybsusc}). We then obtain 
$T_{irr}$ = 17.0\,K. The size of the low field irreversibility 
as evidenced by the difference between the the FC and ZFC branches, 
is quite small and it is somewhat different to that observed in 
Y$_2$Mo$_2$O$_7$ 
\,\cite{greedan86}. The question then arises as to the nature 
of the transition occurring at $T_{irr}$. 
We first note that cases are known where antiferromagnetic (AF) ordering 
takes place without giving rise to an anomaly in the magnetic susceptibility 
(see for instance, Ref.\onlinecite{ott} for the case of the Yb pnictides). 
Thus, the fact that there is no strong anomaly does not in itself represent 
evidence against the existence of low temperature magnetic order or 
correlations. In fact, as is described in section \ref{secmossyb}, at
temperatures below $T_{irr}$, the $^{170}$Yb M\"ossbauer data evidence 
Yb$^{3+}$ magnetic hyperfine fields and Yb$^{3+}$ magnetic moments which are 
produced through Mo-Yb exchange. $T_{irr}$ could then correspond to a change 
in the length scale and/or in the fluctuation rate of the correlations within 
the Mo sublattice.

\begin{figure}
\epsfxsize=7 cm
\centerline{\epsfbox{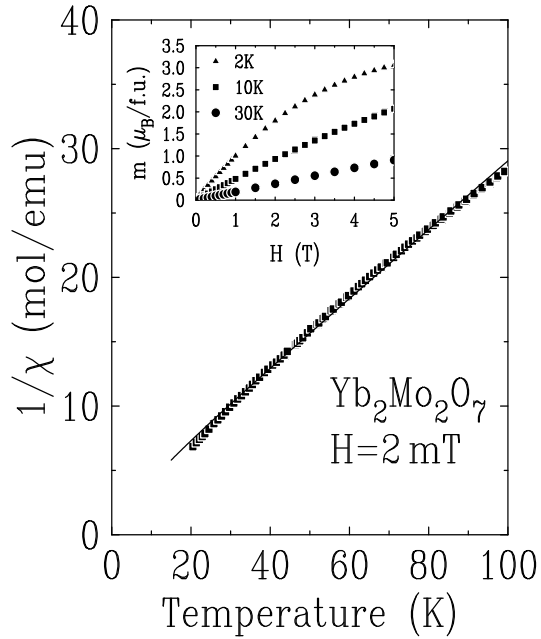}}
\vspace {0.5cm}
\caption{Inverse magnetic susceptibility in Yb$_2$Mo$_2$O$_7$
with a field of 2\,mT; the solid line is a fit as explained in the text. 
Inset: field dependence of the isothermal magnetisation per formula unit
at selected temperatures.}
\label{figybmagn}
\end{figure}

The value of $T_{irr}$ can be linked to one of two possible scenarios. 
First, that frustation does not play a major role so that $T_{irr}$ directly 
corresponds to the strength of the molecular field interaction within the 
Mo$^{4+}$ sublattice. Second, that frustration is operative and in common
with the typical behaviour of frustrated compounds, the temperature of the 
transition ($T_{irr}$) is much smaller than the temperature equivalent to the 
strength of the Mo-Mo coupling. We examine the two possibilities in turn.
For the first case, we analyse the inverse susceptibility 1/$\chi$ between 
20-100\,K (Fig.\ref{figybmagn}) in terms of the molecular field approximation.
In this temperature range, 1/$\chi$ shows a quasi-linear thermal dependence, 
with a small background curvature due to the influence of the three Yb$^{3+}$
Kramers doublet excited crystal field levels which have energies up to
$\sim$ 1000\,K (see section \ref{pac}). Below 100\,K, only the ground crystal 
field Yb$^{3+}$ doublet is appreciably populated and it is described by an 
effective spin 1/2 and a crystal field derived spectroscopic g-tensor. As 
Yb$_2$Mo$_2$O$_7$ is an insulator, an ionic description for the Mo$^{4+}$ ion,
{\it i.e.} S=1 and a g-factor close to 2, should be adequate. 
We assume the following
hierarchy of exchange interactions: a dominant Mo-Mo coupling which is AF 
and a smaller Mo-Yb exchange, with a molecular field constant $\lambda$. 
For a two magnetic sublattice system in the paramagnetic region, the molecular
field theory leads to a ferrimagnetic-like inverse susceptibility:

\begin{equation}
{1 \over \chi} = {{T(T+T_N)-\lambda^2C_{Mo}C_{Yb}} \over
{T(C_{Mo}+C_{Yb})+2\lambda C_{Mo}C_{Yb}+C_{Yb}T_N}},
\label{suscmol}
\end{equation}

\noindent
where $C_{Mo}$ and $C_{Yb}$ are the Curie constants respectively of Mo$^{4+}$
and Yb$^{3+}$. Setting $T_N = T_{irr} = 17$\,K, expression (\ref{suscmol}) 
reproduces the main features of the thermal variation of 1/$\chi$ (solid line 
in Fig.\ref{figybmagn}) with a Mo$^{4+}$ g-value of 1.9(1) (close to the ionic
value appropriate for a high spin Mo$^{4+}$ ion) and an average  Yb$^{3+}$ 
g-value of 3.2(2). This latter value is coherent with the average saturated 
magnetic moment of 1.7\,$\mu_B$ measured by $^{170}$Yb M\"ossbauer 
spectroscopy (see section \ref{secmossyb}). The Mo-Yb exchange constant 
$\lambda$ derived from this analysis is antiferromagnetic, and its value lies 
in the range from $-0.82$ to $-2.20$\,T/$\mu_B$. This yields a mean saturated 
molecular field acting on the Yb$^{3+}$ ion of 2.75\,T  and a mean Yb-Mo 
exchange energy of about 3\,K.
In agreement with this analysis, the direct extrapolation of the inverse
susceptibility (Fig.\ref{figybmagn}) leads to a paramagnetic Curie-Weiss 
temperature with a value in the range \ -5 to -10\,K (AF interaction).
The AF nature of the correlations is confirmed by two aspects of the 
magnetisation measurements which are shown in the inset of Fig.\ref{figybmagn}
: the increase of the moment with applied field is relatively slow 
and at 2\,K, saturation is not achieved in a field of 5\,T.

The possibility that the strength of the Mo - Mo sublattice interaction
is very much bigger than that corresponding to $T_{irr}$ is based, for the 
moment, only on the analogy with the situation in Y$_2$Mo$_2$O$_7$. In this
case, muon spin relaxation ($\mu$SR) 
\cite{dunsiger96}, neutron diffraction 
\cite{gardner99} and unpublished high temperature susceptibility
measurements quoted in Ref. 
\onlinecite{gardner99} all point to the presence of a Mo - Mo interaction 
which is much bigger than that corresponding to $T_{irr}$. It is not feasible 
to access the Mo - Mo interaction in Yb$_2$Mo$_2$O$_7$ from high temperature 
susceptibililty measurements because of the temperature dependent
contribution coming from the Yb$^{3+}$ excited crystal field levels. To our 
knowledge, no $\mu$SR or neutron diffraction measurements have yet been 
carried out on Yb$_2$Mo$_2$O$_7$.

\subsection {$^{170}$Yb M\"ossbauer measurements.}
\label{secmossyb}

The $^{170}$Yb M\"ossbauer absorption measurements were made over the 
temperature range 95 to 0.036\,K using a source of Tm$^{\ast}$B$_{12}$
and a triangular velocity sweep. For $^{170}$Yb, the ground nuclear state has 
a spin I$_g$ = 0, the excited nuclear state has a spin I$_{ex}$ = 2 and a
quadrupole moment $Q=-2.11$\,b, and E$_{\gamma}$ = 84.3\,keV and 1\,cm/s 
corresponds to 680\,MHz. 

\begin{figure}
\epsfxsize=6 cm
\centerline{\epsfbox{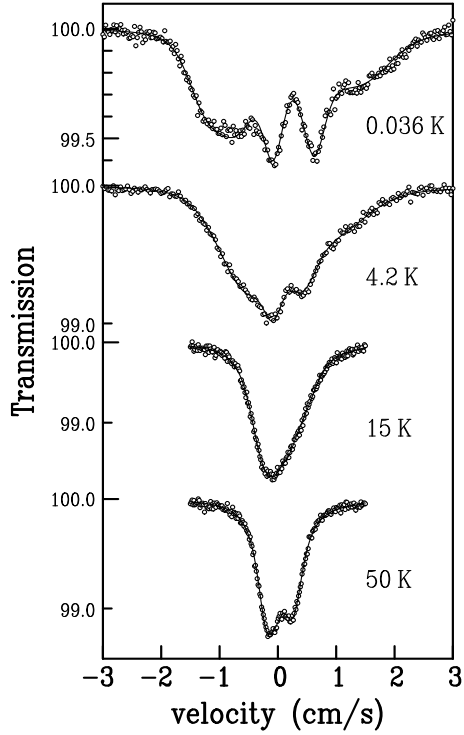}}
\vspace {0.5cm}
\caption{$^{170}$Yb~ M\"ossbauer~ absorption~ spectra for Yb$_2$Mo$_2$O$_7$. 
At 50\,K, only a quadrupole hyperfine interaction is present. To obtain the
satisfactory linefit shown, it is necessary to allow this interaction to show
a distribution and to introduce an asymmetry parameter. At the three lower 
temperatures an additional magnetic hyperfine interaction is also present.}
\label{figybmoss}
\end{figure}

In the upper part of the examined temperature range, the spectra 
are well described in terms of quadrupole hyperfine interactions, showing that
in this region, the Yb$^{3+}$ ions are paramagnetic. As exemplified by the 
data at 50\,K on Fig. \ref{figybmoss}, we find that an acceptable 
data fit cannot be obtained if the quadrupole hyperfine interaction is taken
to have the axial symmetry that is characteristic of the rare earth site
in the standard pyrochlore structure. To obtain a good quality data fit, it 
is necessary both to introduce an asymmetry parameter ($\eta$) and to allow 
the principal component $V_{ZZ}$ of the electric field gradient tensor to 
show a distribution. We find  $\eta$ = 0.6 and with a Gaussian shaped 
distribution, we obtain a mean value: eQV$_{ZZ}$/8 = 1.3\,mm/s and a root mean
square deviation: $\sigma \simeq$ 0.4\,mm/s.
The microscopic M\"ossbauer probe measurements thus identify two aspects
of the local structure of Yb$_2$Mo$_2$O$_7$: the point symmetry at the Yb 
site is non-axial and there is a distribution of environments 
indicating that significant disorder is present.
Lattice disorder has previously been evidenced in the isomorphous 
semiconducting compound Y$_2$Mo$_2$O$_7$ by X-ray absorption edge 
\cite{booth00} and $^{89}$Y nuclear magnetic resonance 
\cite{keren01} 
measurements. There was no evidence of any significant departure from full 
oxygen stoichiometry nor of site exchange and it was suggested that the 
disorder principally involved the Mo-Mo pair distances with also some 
modest disorder in the Y-O(1) distances
\cite{booth00}.
It seems likely that an analogous situation will also pertain in 
Yb$_2$Mo$_2$O$_7$ and that the anomalous characteristics of the
quadrupole hyperfine interaction reported here are linked to lattice disorder
involving some of the bond lengths and angles.

At low temperatures, magnetic hyperfine splittings are visible
(Fig.\ref{figybmoss}). Such splittings appear when the Yb$^{3+}$
moments are either long or short range magnetically correlated and when any 
fluctuation of the correlated moments occurs at frequencies of the order of or
below the 
$^{170}$Yb M\"ossbauer threshold value $\sim$ 3.0 $\times $10$^{8}$ s$^{-1}$. 
We first consider the data at 0.036\,K. The overall shape of the spectrum and 
in particular the inhomogeneous broadenings of the 5 lines, point to the 
presence of a distribution in hyperfine field and quadrupolar coupling 
parameter values. This is coherent with the analysis of the data in the 
paramagnetic region, which evidenced a distribution of quadrupole hyperfine 
interactions. 
Both distributions can be linked to distribution in the Yb$^{3+}$ 
ground state wave functions which results from the local disorder.
In presence of random disorder, it can be shown that the distributions
in the hyperfine field and quadrupolar parameters are linearly 
correlated 
\cite{ybcual}. 
Using linearly correlated distributions, we obtain good data fits
(Fig.\ref{figybmoss}) and these provide the mean $^{170}$Yb hyperfine field 
and the root mean square deviation of the distribution. The saturated values 
at 0.036\,K are respectively 175\,T and 45\,T. The mean saturated Yb$^{3+}$ 
magnetic moment value is therefore 1.7\,$\mu_B$, using the standard 
$^{170}$Yb$^{3+}$ relation: 1$\mu_B$ corresponds to 102\,T.

\begin{figure}
\epsfxsize=7 cm
\centerline{\epsfbox{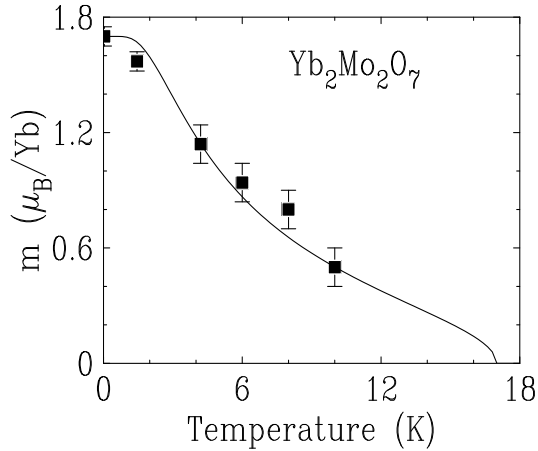}}
\vspace{0.5cm}
\caption{Thermal dependence of the mean Yb$^{3+}$ magnetic moment in 
Yb$_2$Mo$_2$O$_7$ obtained from $^{170}$Yb 
M\"ossbauer measurements. The solid line is a fit to a molecular field model 
as explained in the text.}
\label{figybchvst}
\end{figure}

The spectra become less resolved as the temperature increases
(Fig.\ref{figybmoss}). Reliable fits in terms of a distribution of hyperfine 
fields are possible only up to 10\,K. The thermal variation of the mean 
Yb$^{3+}$ moment is shown on Fig.\ref{figybchvst}. The theoretical curve is 
obtained assuming that the
Yb$^{3+}$ ground doublet, with effective spin 1/2 and spectroscopic factor g,
is polarised by the molecular field $H_{ex}$ due to Mo-Yb exchange. 
In order to calculate H$_{ex}$(T), which is proportional to the Mo moment
within a molecular field model, we assume that the thermal variation of 
the Mo moment follows the mean field law for S=1. Then the reduced exchange 
field
$\sigma(T) = H_{ex}(T)/H_{ex}(0)$
is obtained by solving the self-consistent equation:
\begin{eqnarray}
 \sigma(T) = {\cal B}_1 
\bigg( {{3 S} \over {S + 1}} {\sigma(T) \over \tau} \bigg),
\label{eqnhffieldb}
\end{eqnarray}
where ${\cal B}_{1}$ is the Brillouin function for S = 1 and $\tau =T/T_N$.
The solid line in Fig.\ref{figybchvst} is obtained with T$_N$ = 17\,K and 
$H_{ex}(0)=3$\,T, corresponding to an exchange energy of 3.4\,K. 
This value is in good agreement with that derived from the analysis of the 
susceptibility data (section \ref{secsusc}).

Although no reliable values can be obtained for the hyperfine field as the 
temperature is increased above $\sim$ 10\,K, it is clear that correlated 
Yb$^{3+}$ magnetic moments are still present. In fact, the lineshapes also
show that some correlations persists above 17\,K. Both just above and 
just below the irreversibility temperature, the M\"ossbauer line shapes
are compatible with the presence of dynamic magnetic correlations.

\subsection {$^{172}$Yb perturbed angular correlation (PAC) measurements.}
\label{pac}

\begin{figure}
\epsfxsize=7 cm
\centerline{\epsfbox{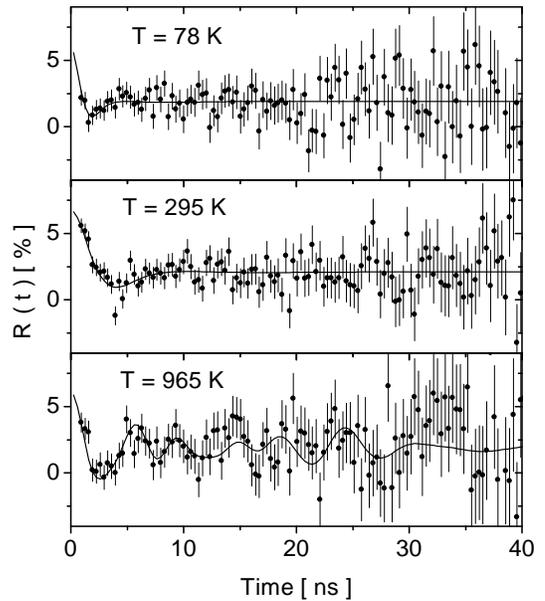}}
\vspace{0.5 cm}
\caption{$^{172}$Yb perturbed angular correlation spectra for 
Yb$_2$Mo$_2$O$_7$.
The solid lines show the fitted perturbation factor due to a distribution of
electric quadrupolar interactions in the $I=3$ intermediate level.}
\label{figybpacspectra}
\end{figure}

The PAC measurements provide the thermal dependence of the absolute value of
the quadrupole hyperfine interaction at the $^{172}$Yb nucleus 
\cite{hodges01,krolas96}.
They were made over the range 34 to 965\,K, that is at temperatures above the 
range where magnetic correlations are present. Three examples of the 
spectra are shown in Fig. \ref{figybpacspectra}.

The strong damping of the $R(t)$ oscillations visible in 
Fig. \ref{figybpacspectra}
is very characteristic of a hyperfine interaction which is distributed in
size. The spectra cannot be fitted in terms of a single quadrupole interaction
at any temperature and it is necessary to allow for a distribution of the 
fitted electric field gradient. This observation reinforces
the same conclusion obtained from the $^{170}$Yb M\"ossbauer analysis.

\begin{figure}
\epsfxsize=7 cm
\centerline{\epsfbox{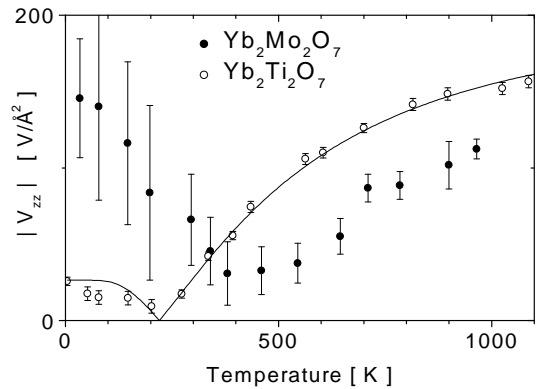}}
\vspace{0.5 cm}
\caption{Full points: thermal dependence of the absolute size of the principal
component $V_{ZZ}$ of the electric field gradient tensor in 
Yb$_2$Mo$_2$O$_7$ obtained from $^{172}$Yb perturbed angular correlation 
measurements. At all temperatures $V_{ZZ}$ shows a distribution which is more 
important at low temperatures than at high temperature, the
root mean squares of the Gaussian are shown by the vertical bars.
Open points: corresponding results (from Ref.\protect\onlinecite{hodges01})
obtained in Yb$_2$Ti$_2$O$_7$ where essentially no distribution is observed
and where the vertical bars correspond to the experimental error; 
the solid line was calculated using a crystal field model.}
\label{figybpacvzzfnt}
\end{figure}

It turns out that in presence of this distribution, the PAC data are not very 
sensitive to the asymmetry parameter $\eta$. The PAC technique thus cannot 
provide independent evidence confirming the conclusion of the $^{170}$Yb 
M\"ossbauer analysis that this parameter is non-zero.
The thermal variation of the mean absolute value of $V_{ZZ}$ and of $\sigma$
the width of its Gaussian distribution are shown in 
Fig. \ref{figybpacvzzfnt}. The total interaction is made up of two parts, one 
which is temperature dependent arising from the Yb$^{3+}$ 4f shell,
and one which is essentially temperature independent arising from the lattice 
charges. At low temperatures, the 4f electron contribution dominates
whereas at high temperatures, the total 4f contribution tends to zero and the
lattice contribution dominates.

At the lowest measurement temperature, the mean value for $\vert V_{ZZ} \vert$
agrees with that obtained from the $^{170}$Yb M\"ossbauer analysis.
This latter technique also provides the sign of the gradient showing that
in the low temperature limit  $V_{ZZ}$ = -135 V/$\AA^2$. 
The two techniques also provide essentially the same value for $\sigma$, the 
root mean square deviation of the distribution.

As shown on Fig.\ref{figybpacvzzfnt}, the electric field gradient also 
evidences a distribution in the high temperature region of the measurement 
range. Since the electric field gradient in this region is due chiefly to the 
surrounding lattice charges, the observation of a distribution directly 
evidences the presence of lattice disorder. We attribute the fact that the 
size of the distribution of the electric field gradient is much bigger at low 
temperatures to the fact that the disorder gives rise to a distribution in 
the Yb$^{3+}$ crystal field parameters and wave functions and this leads to 
an enhanced distribution in the part of the quadrupolar hyperfine interaction 
linked with the 4f-shell.

Because of the local symmetry lowering, it is impossible to carry out a 
crystal field analysis of the thermal variation of 
$V_{ZZ}$, such as was done for Yb$_2$Ti$_2$O$_7$ 
\,\cite{hodges01}. The 
values for this latter compound, also shown in 
Fig.\ref{figybpacvzzfnt}, are different from those in Yb$_2$Mo$_2$O$_7$ 
showing there are some differences between the Yb$^{3+}$ crystal field 
properties of the two compounds. The temperature range over which the 
gradient continues to vary is however similar in the two cases suggesting the 
overall crystal field splittings are of comparable magnitudes ($\sim$ 1000\,K)
in the two compounds.

\section {Gd$_2$Mo$_2$O$_7$.}
\label{sectiongd}

\subsection{Susceptibility and magnetisation measurements.}
\label{suscgd}

\begin{figure}
\epsfxsize=7 cm
\centerline{\epsfbox{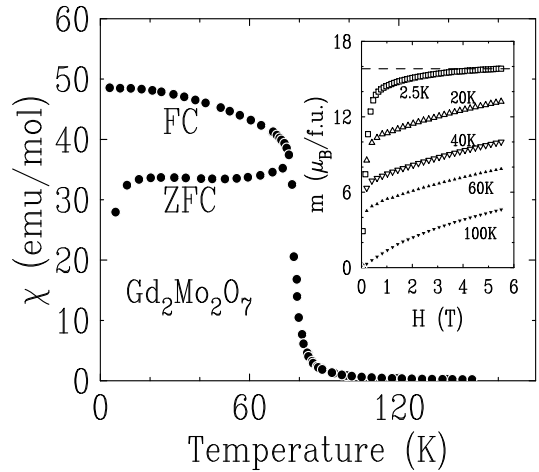}}
\vspace {0.5cm}
\caption{Magnetic susceptibility of Gd$_2$Mo$_2$O$_7$ with an applied field of
0.2\,mT. FC: field cooled and  ZFC: zero field cooled.
{\bf Inset}: isothermal magnetisation curves at selected temperatures.}
\label{figgdsusc}
\end{figure}

The magnetic susceptibility $\chi(T)$ and magnetisation $m(H)$ measurements 
are shown in Fig.\ref{figgdsusc}.
The sharp rise in $\chi(T)$ that occurs at $\sim$ 80\,K (previously observed
at $\sim$ 55\,K 
\,\cite{ali89,raju92} and at $\sim$ 65\,K 
\,\cite{katsufuji00}) 
suggests the onset of ferromagnetic ordering of the Mo$^{4+}$ moments. 
The specific heat shows only a broad anomaly near the temperature of the 
susceptibility rise, 
\cite{raju92}, suggesting the ordering is short range.
An irreversibility between the FC and ZFC branches (reported previously
\cite{ali89,raju92}) is observed below 75\,K. It is probably chiefly related
to the pinning of domain walls by defects.  

The rapid initial rise of the magnetisation curves (inset of 
Fig.\ref{figgdsusc}) measured below 60\,K also evidences ferromagnetic 
interactions.
At 2.5\,K, the quasi-saturated value of the high field magnetisation 
of 15.8(4)\,$\mu_B$ per formula unit (dashed line in the inset of 
Fig.\ref{figgdsusc}) shows the Gd-Mo coupling is ferromagnetic. The two 
Gd$^{3+}$ ions contribute a saturated moment of 14\,$\mu_B$, and thus each 
Mo$^{4+}$ ion carries a moment of 0.9(2)\,$\mu_B$, parallel to that of the
Gd$^{3+}$, as reported previously 
\cite{greedanLB92}. This value for the 
Mo$^{4+}$ moment is much smaller than that expected 
assuming a ionic description in terms of high spin 4d$^2$, t$_{2g}$ state 
with S=1 and g$\sim$2 ($\sim2\mu_B$). 
Furthermore, our attempts to fit the magnetisation curves within a 
self-consistent molecular field model assuming an ionic description for 
Mo$^{4+}$ as well as for Gd$^{3+}$ failed even allowing the Mo g-factor 
to be adjustable. The magnetic properties of the Mo in metallic 
Gd$_2$Mo$_2$O$_7$ thus appear to be associated rather with exchange coupled 
4d electrons with some metallic character.

\subsection{$^{155}$Gd M\"ossbauer measurements.}
\label{mossgd}

\begin{figure}
\epsfxsize=5 cm
\centerline{\epsfbox{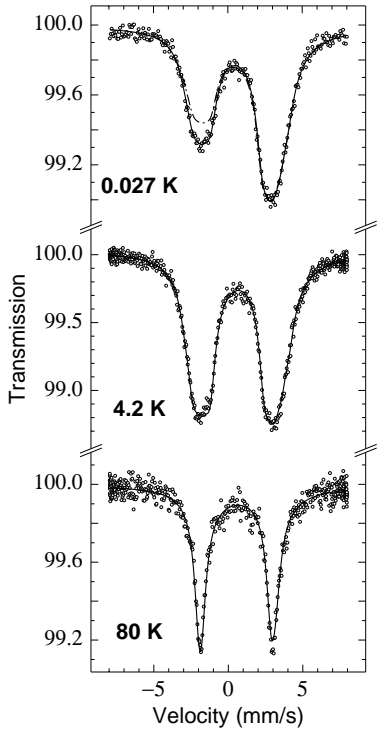}}
\vspace {0.5cm}
\caption{$^{155}$Gd~M\"ossbauer~absorption~spectra for Gd$_2$Mo$_2$O$_7$.
The two data fits (full and dashed lines) 
shown at 27\,mK are discussed in the text in relation to the presence
of out of equilibrium hyperfine level populations.}
\label{figgdmoss}
\end{figure}

The $^{155}$Gd M\"ossbauer absorption measurements were made over the 
temperature range 80 to 0.027\,K using a source of Sm$^{\ast}$Pd$_{3}$. 
The isotope $^{155}$Gd has a ground nuclear state with
spin I$_g$ = 3/2 and quadrupole moment Q=1.31\,barn, and an excited nuclear
state with spin I$_{ex}$ = 5/2 and very small quadrupole moment. The 
transition energy is E$_{\gamma}$ = 86.5\,keV, and 1\,mm/s corresponds 
to 69.8\,MHz. Spectra at some selected temperatures are shown in 
Fig. \ref{figgdmoss}.

At 80\,K, in the paramagnetic phase, the absorption 
takes the form of a nearly symmetric doublet. It corresponds to a quadrupole
hyperfine interaction, with a quadrupolar splitting 
${eQV_{ZZ} \over 2} {(1 + {\eta^2 \over 3})}^{1/2} \simeq - 5.1$\,mm/s 
(the negative sign is obtained from the analysis of the line shapes at low
temperatures when a hyperfine field is also present and $\eta$ is included to
allow for possible local symmetry lowering).
For the S-state Gd$^{3+}$ ion, the quadrupole hyperfine interaction is due 
only to the neighbouring lattice charges. As in other Gd pyrochlores
\cite{cashion73}, the interaction is quite large in keeping with the 
important local structural anisotropy of the rare earth site. 

A hyperfine field, proportional to the Gd$^{3+}$ moment, appears when the 
temperature is lowered below $\sim$ 75\,K, and this confirms the magnetic 
transition evidenced by the susceptibility data. 
In Gd$_2$Mo$_2$O$_7$, the magnetic hyperfine interaction is much 
smaller than the quadrupole interaction, and the hyperfine field acts only to 
broaden each of the two absorption lines. At 0.027\,K, the saturated hyperfine
field is: $H_{hf}(0) \simeq$19.8\,T, which is markedly smaller than that
usually found in insulating Gd compounds ($\sim$30\,T). The reduced value of
$H_{hf}(0)$ in this metallic compound is probably linked to the exchange 
polarisation of $s$-type conduction electrons, which contributes a hyperfine 
field opposite to that arising from the polarisation of the core $s$-electrons
by the 4f shell moment.
At 0.027\,K, the relative intensity of the two absorption lines is different
from that at 4.2\,K even though the value of the hyperfine field remains
essentially the same. The change in the relative intensity at very low 
temperatures is related to changes in the populations of the sublevels of the 
$^{155}$Gd nuclear ground state and these will be discussed at more length in 
section \ref{mossgdex}.

The asymmetry parameter of the electric field gradient tensor ($\eta$) can 
only be obtained from a $^{155}$Gd M\"ossbauer measurement when a magnetic 
hyperfine interaction is present. From the  0.027\,K spectrum, we obtain 
$0.0 \le \eta \le 0.4$. Thus, although it is possible that $\eta$ is
zero as expected for the axially symmetric rare earth site of the standard
pyrochlore structure, we cannot discount the possibility (clearly established
above for Yb$_2$Mo$_2$O$_7$) that $\eta$ is not exactly zero in which case 
local symmetry lowering is present.
Over the whole temperature range, {\it i.e.}, in both the paramagnetic and 
magnetically correlated regions, we find the line widths of the individual 
transitions are $\sim$ 20\% larger than in the compound Gd$_2$Sn$_2$O$_7$ 
where $\eta$ is essentially zero 
\cite{bertin02}. We attribute the background line broadening in 
Gd$_2$Mo$_2$O$_7$ to indicate the quadrupolar hyperfine interaction shows a 
small distribution. Since this interaction arises solely from the electric 
field gradient created by the lattice charges, the presence of a distribution
directly provides evidence for local disorder.

In principle, the analysis of the M\"ossbauer absorption provides the 
angle $\theta$ between the direction of the magnetic moment and the principal 
local axis of the electric field gradient tensor (a [111] direction). The best
fit yielded $\theta \simeq $55$^\circ$ but only marginally poorer quality fits
were obtained if $\theta$ was assumed to be distributed at random. The reduced
sensitivity of the data fits to the direction of the hyperfine field is linked
to the fact that the magnetic hyperfine interaction is much smaller than the 
quadrupole hyperfine interaction. 
Since the Gd$^{3+}$ ion is essentially isotropic, the direction of the 
hyperfine field corresponds to the direction of the exchange field coming from
the Mo$^{4+}$ sublattice. The choice  of a unique angle 
$\theta \simeq $55$^\circ$ is compatible with a magnetic structure where the 
Gd$^{3+}$ and the Mo$^{4+}$ moments are both aligned close to a [100] axis.
To our knowledge, Gd$_2$Mo$_2$O$_7$ has not been studied by neutron 
diffraction. Such measurements are, in fact,  difficult due to the high 
neutron absorption cross section of non-enriched Gd. They would be useful in
order to check if the moments are aligned towards a [100] direction and also, 
in fact, to establish whether the magnetic correlations are short or long 
range. We recall that a specific heat analysis in  Gd$_2$Mo$_2$O$_7$ suggested
that there is no long range order
\cite{raju92} and that in in Nd$_2$Mo$_2$O$_7$, which is also metallic, 
neutron diffraction measurements showed the magnetic order is long range 
\cite{yasui01} and the ferromagnetically coupled Mo$^{4+}$ moments are 
aligned close (from 6 to 9$^{\circ}$) to a [100] direction.

In their analysis of the magnetic specific heat in Gd$_2$Mo$_2$O$_7$, 
the authors of Ref.\onlinecite{raju92} assume that even at 
very low temperatures, the exchange field experienced by the Gd$^{3+}$ ion is 
distributed in size, with a finite weight at zero field. We tried to check 
this assumption by simulating spectra with such a distribution but no 
clear conclusion could be reached.

\begin{figure}
\epsfxsize=7 cm
\centerline{\epsfbox{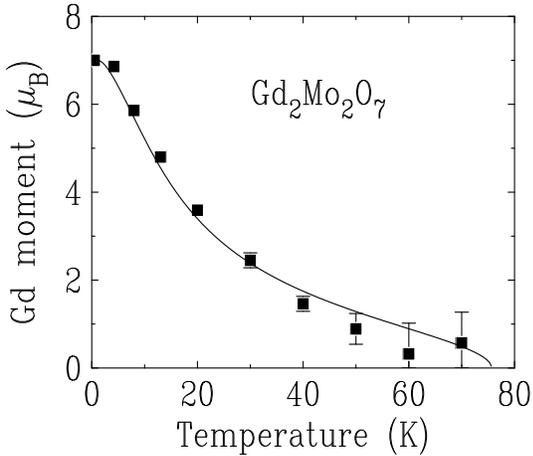}}
\vspace {0.5 cm}
\caption{Thermal variation of the
Gd$^{3+}$ 4f shell magnetic moment in Gd$_2$Mo$_2$O$_7$ obtained from
$^{155}$Gd M\"ossbauer measurements.
The solid line is a fit with the molecular field model described in the text.}
\label{figgdmomvst}
\end{figure}

Fig.\ref{figgdmomvst} shows the thermal variation of the Gd$^{3+}$ 
4f shell magnetic moment, obtained from the hyperfine field value by scaling, 
using the relation that the measured saturated hyperfine field of 19.8\,T 
corresponds to a saturated Gd$^{3+}$ moment of 7\,$\mu_B$.
In order to analyse these data, we assume the usual hierarchy of exchange 
interactions, in decreasing order: Mo-Mo, responsible for the
transition at 75\,K, Mo-Gd, responsible for the Gd$^{3+}$ polarisation below
75\,K, and Gd-Gd, which we will neglect here. 
Then, the Gd$^{3+}$ magnetic moment at a temperature T is given by:
\begin{eqnarray}
M_{Gd}(T)  = M_{Gd}(0) \ {\cal B}_{7/2} 
\bigg( {{M_{Gd}(0) H_{ex}(T)} \over {k_B T}}\bigg),
\label{eqnhffielda}
\end{eqnarray}
where ${\cal B}_{7/2}$ is the Brillouin function for S = 7/2 
and H$_{ex}$(T) is the Mo derived exchange field which is obtained
by solving Eqn.\ref{eqnhffieldb} with $\tau =T/T_C$ and $T_C=75$\,K, 
the Mo$^{4+}$ ordering temperature.
The solid line on Fig.\ref{figgdmomvst}, which provides a good fit to the 
experimental data, is obtained with a Mo-Gd exchange field 
H$_{ex}$(0) = 5.5\,T. 
The Mo-Gd exchange energy can then be estimated
as: $E_{ex} = M_{Gd}(0) H_{ex}(0) \simeq 26$\,K. This is smaller than $T_C$
and so {\it a posteriori} justifies the approximations made in the above
calculation. The mean size of the Mo$^{4+}$ derived field acting on the 
Gd$^{3+}$ obtained here is $\sim 30\%$ lower than that estimated from a
specific heat analysis 
\cite{raju92}.
The fact that a mean field law is able to account for
the thermal variation of the Gd moment, whereas it fails to reproduce the
magnetisation curves (Fig.\ref{figgdsusc}, inset), probably stems from the 
fact that M$_{Gd}$(T) is not very sensitive to the exact shape of H$_{ex}$(T),
and that for Gd$^{3+}$ an ionic description is valid.

\subsection{$^{155}$Gd M\"ossbauer evidence for persisting T $\to$ 0 magnetic 
fluctuations.}
\label{mossgdex}

The evidence for the persistence of the magnetic fluctuations is based on the
analysis of the relative intensities of the M\"ossbauer absorption lines 
of the 27\,mK data on Fig.\ref{figgdmoss}. We have described this approach
previously 
\cite{bertin02} for the case of the pyrochlore Gd$_2$Sn$_2$O$_7$ and recall 
the main features.
For the values of the saturated magnetic and quadrupole hyperfine interactions
established above, the four hyperfine levels of the $^{155}$Gd I = 3/2 ground 
state span an energy range of 20\,mK. Below $\sim$ 100\,mK, the relative 
populations of these levels will differ from the equipopulated values that 
pertain at higher temperatures and this leads to changes in the relative
intensities of the absorption lines. By measuring these relative intensities 
on the low temperature spectrum, it is possible to obtain the 
effective temperature of the hyperfine levels, which in conventional
magnetically ordered magnetic compounds corresponds to the temperature of the 
sample.
We find an effective hyperfine level temperature of 47(8)\,mK 
(full line fit on Fig.\ref{figgdmoss}) which is significantly higher 
than the measurement probe temperature (27\,mK), {\it i.e} the lattice 
temperature of the sample (dashed line on Fig.\ref{figgdmoss}). 
In other words, the steady state populations of the hyperfine levels are not 
those corresponding to thermal equilibrium. This indicates that at 27\,mK, 
there is a finite Gd$^{3+}$ spin flip time which is of the same magnitude as 
the nuclear relaxation time.
This approach cannot provide quantitative information concerning the 
rate of the spin fluctuations since the nuclear relaxation time is unknown. 
We recall that the good quality data fits on Fig.\ref{figgdmoss} were made 
with the assumption of a ``static'' hyperfine field, {\it i.e}. a hyperfine 
field (or Gd$^{3+}$ magnetic moment) which appears static on the scale of the 
$^{155}$Gd hyperfine Larmor frequency 
($\sim$1.2$\times 10^8$\,s$^{-1}$). The low temperature Gd$^{3+}$ spin 
fluctuations that are evidenced by the anomalous hyperfine level populations 
thus occur with frequencies below this value. Since the fluctuations of the 
Gd$^{3+}$ moments are driven by the fluctuations of the Mo$^{4+}$ derived 
exchange field, this shows the correlated moments of the Mo$^{4+}$ sublattice 
also continue to fluctuate at 27\,mK.

\section{Summary and Discussion.}
\label{sectiondisc}

We have studied the pyrochlore compounds Yb$_2$Mo$_2$O$_7$ and 
Gd$_2$Mo$_2$O$_7$ by microscopic hyperfine techniques ($^{170}$Yb, 
$^{155}$Gd M\"ossbauer and $^{172}$Yb perturbed angular correlation
spectroscopies) and bulk magnetic measurements and we have obtained 
information concerning three aspects of their properties : crystallographic 
disorder, the magnetic interactions and the role of magnetic frustration.  

In Yb$_2$Mo$_2$O$_7$, the Yb$^{3+}$ site symmetry is lower than that of the 
rare earth site in the standard pyrochlore structure. The distortions are 
local, since the room temperature X-ray diffraction spectra indicate that the 
overall cubic lattice symmetry is preserved. In addition, there is 
considerable lattice disorder. In Gd$_2$Mo$_2$O$_7$, there is no definite 
evidence of local symmetry lowering but this cannot be excluded and there is 
some evidence of local disorder. 

In semi-conducting Yb$_2$Mo$_2$O$_7$, the Mo-Mo exchange coupling is dominated
by an interaction which is antiferromagnetic as is the case in semi-conducting
Y$_2$Mo$_2$O$_7$.
Based on the analysis of the magnetic susceptibility in the low temperature 
paramagnetic phase within a molecular field model, the strength of this
coupling is of the same order of magnitude as the irreversibility temperature 
($T_{irr}$ = 17\,K). In this case, the Mo-Mo exchange in Yb$_2$Mo$_2$O$_7$ is
much smaller than that in Y$_2$Mo$_2$O$_7$. It is possible however, that the 
Mo-Mo coupling is much bigger than than that correponding to $T_{irr}$.
Additional studies (for example, in field $^{57}$Fe M\"ossbauer, $\mu$SR or 
neutron diffraction measurements) are needed to investigate this point.
At low temperatures, the Yb$^{3+}$ are magnetically polarised by the field
coming from the Mo$^{4+}$ sublattice. The mean value of the saturated 
field reaching a Yb$^{3+}$ is 3\,T and the mean value of the saturated
Yb$^{3+}$ moment is 1.7$\mu_B$.
In metallic Gd$_2$Mo$_2$O$_7$, ferromagnetic correlations dominate, as shown 
by the saturation of the magnetisation at 2\,K and
by the sharp rise in the susceptibility at 80\,K. Below this latter 
temperature, the Gd$^{3+}$ are magnetically polarised by the field coming
from the Mo$^{4+}$ sublattice and the saturated value of this field is 
5.5\,T.

In Y$_2$Mo$_2$O$_7$, Mo spin fluctuations are known to persist as T $\to$ 0
\,\cite{dunsiger96,gardner99}. It seems likely that such fluctuations will
also be present in Yb$_2$Mo$_2$O$_7$.
Because the hyperfine fields in  Yb$_2$Mo$_2$O$_7$
appear static on the $^{170}$Yb M\"ossbauer frequency scale of 
$\sim$ 3.0 $\times$ 10$^8$\,s$^{-1}$, any fluctuations that occur must take 
place at frequencies that are lower than this value.
In Gd$_2$Mo$_2$O$_7$, we find that Gd$^{3+}$ and Mo$^{4+}$ spin 
fluctuations persist as T $\to$ 0 so evidencing spin liquid behaviour.
The continued influence of frustration is somewhat surprising for it 
concerns a case where the coupling is ferromagnetic and where no important 
single ion anisotropy is involved.  
It is possible that the frustration is linked to the role of 
further neighbour interactions which are antiferromagnetic. 

For both a semi-conducting pyrochlore where the dominant Mo-Mo interaction is 
anti-ferromagnetic (Y$_2$Mo$_2$O$_7$
\,\cite{dunsiger96,gardner99}) and a metallic pyrochlore (Gd$_2$Mo$_2$O$_7$, 
present results) where the dominant interaction is ferromagnetic, the spin 
fluctuations persist as 
T$\to$ 0.

\widetext
\end{document}